\documentclass{article}
\usepackage[utf8]{inputenc}
\usepackage{authblk}
\usepackage{setspace}
\usepackage[margin=1.25in]{geometry}
\usepackage{graphicx}
\graphicspath{ {./figures/} }
\usepackage{subcaption}
\usepackage{amsmath}
\usepackage{lineno}
\usepackage{units}
\usepackage{changes}
\DeclareUnicodeCharacter{0301}{\'{e}}

\usepackage[style=nejm, 
citestyle=numeric-comp,
sorting=none]{biblatex}
\title{Spatial aberrations in high-order harmonic generation}
\author[1,$\dagger$,*]{Marius Plach}
\author[1,2,3,$\dagger$,*]{Federico Vismarra}
\author[1]{Elisa Appi}
\author[1]{Vénus Poulain}
\author[1]{Jasper Peschel}
\author[4]{Peter Smorenburg}
\author[4]{David P. O'Dwyer}
\author[4]{Stephen Edward}
\author[4]{Yin Tao}
\author[3]{Rocío Borrego-Varillas}
\author[3]{Mauro Nisoli}
\author[1]{Cord L. Arnold}
\author[1]{Anne L'Huillier}
\author[1]{Per Eng-Johnsson}

\affil[1]{Department of Physics, Lund University, P.O. Box 118, 22100 Lund, Sweden.}
\affil[2]{Department of Physics, Politecnico di Milano, 20133 Milano, Italy.}
\affil[3]{Institute for Photonics and Nanotechnologies, IFN-CNR, 20133 Milano, Italy.}
\affil[4]{ASML Research, ASML Netherlands B.V., 5504 DR Veldhoven, Netherlands.}

\affil[*]{Address correspondence to: marius.plach@fysik.lth.se, federico.vismarra@polimi.it}
\affil[$\dagger$]{These authors contributed equally to this work.}

\begin{document}
\maketitle

\begin{abstract}
We investigate the spatial characteristics of high-order harmonic radiation generated in argon, and observe cross-like patterns in the far field. An analytical model describing harmonics from an astigmatic driving beam reveals that these patterns result from the order and generation position dependent divergence of harmonics. Even small amounts of driving field astigmatism may result in cross-like patterns, coming from the superposition of individual harmonics with spatial profiles elongated in different directions. By correcting the aberrations using a deformable mirror, we show that fine-tuning the driving wavefront is essential for optimal spatial quality of the harmonics.
\end{abstract}


\section{Introduction}
\begin{figure}[b]\centering
\hspace*{-0.5cm}
\includegraphics[width=14 cm]{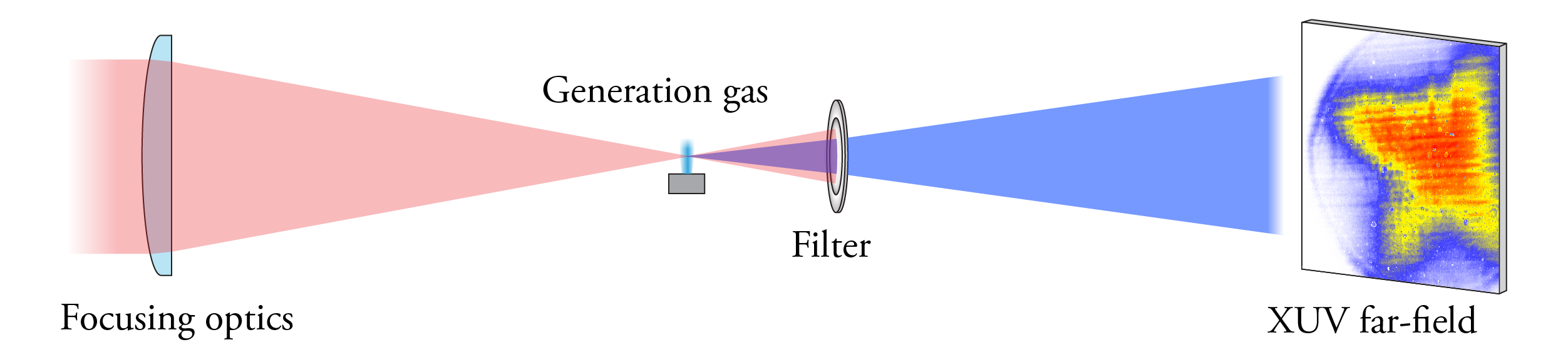}
\hspace*{+0.5cm}
\caption{Illustration of the effect discussed in the present work. A strong IR field is focused in a gas target and generates high-order harmonics. After filtering out the driving field, the spectrally-integrated XUV far-field is often observed to be strongly distorted, presenting a typical cross-like pattern, even for seemingly aberration-free driving fields.}
\label{fig:Figure_1_schematic}
\end{figure}
High-order harmonic generation (HHG) allows extending the spectral range accessible with conventional laser technology into the extreme ultraviolet (XUV) regime, and decreasing the pulse duration into the attosecond temporal regime. Due to their high temporal and spatial coherence, HHG sources are essential tools in many research fields, from the study of ultrafast processes in atoms, molecules and solids~\cite{RevModPhys.81.163,BorregoVarillas2022} to coherent diffraction imaging~\cite{Zrch2014}, nano-scale metrology~\cite{sakdinawat2010nanoscale}, ptychography~\cite{Loetgering:22}, and more.
As shown schematically in Fig.~\ref{fig:Figure_1_schematic}, an HHG setup requires a driving femtosecond laser field focused to high intensity in a generation medium. Typically, near-infrared (IR) or mid-IR laser pulses are employed to generate high-order harmonics in the XUV or soft X-ray spectral range by interaction with a gas in a cell, jet, or capillary~\cite{popmintchev2012bright}.

The generation conditions (including the laser wavelength, laser intensity profile, laser wavefront, gas medium density, atomic species, and the position of the generation medium with respect to the laser focus) highly impact the properties of the generated radiation. Therefore, the characterization and control of those conditions are of critical importance for HHG optimization, not only in terms of the generated photon flux, but also for the spatial quality of the XUV radiation. {The presence of significantly distorted XUV wavefronts can give rise to deformed far-field spatial profiles, exemplified by the cross-like patterns shown in Figure ~\ref{fig:Figure_1_schematic}. These distortions impose severe limitations on using XUV profiles in applications demanding high spatial-quality beams.}
Other studies have noted similar far-field profiles {without directly addressing their origin}~\cite{Kretschmar:20, Wodzinski:20}. {However, an extended comprehension and control of the XUV wavefront properties is of paramount importance} for high-quality focusing of the XUV beam where a high field strength is required, {e.g. in table-top XUV-pump-XUV-probe experiments~\cite{Makos2020,kretschmar2023compact}}, or for applications where the focus should be contained in a small target, {e.g. nano-scale imaging~\cite{Gardner2017}}. So far, most theoretical studies assume ideal driving fields, and only recently the role of more realistic beams has been addressed~\cite{PhysRevA.95.063823,Milosevic:22,Carlstrm2016,Ensemble2022}. 
Previous studies emphasized the dependence of the XUV wavefront properties on the harmonic order~\cite{frumker2012order}, on the position of the generating medium relative to the laser focus~\cite{HampusSpatioTemporal}, and on the influence of the wavefront and intensity profile of the driving field~\cite{2004OptimizationwithDM, PhysRevA.95.063823, ValentinJOSAB2008, dacasa2019single, Wodzinski:20, Veyrinas2023}. Different harmonics can show very different divergence properties, leading to strong chromatic aberrations after refocusing the full HHG spectrum, and resulting in spatiotemporal coupling in the attosecond pulses~\cite{HampusSpatioTemporal,QuintardSciAdv2019, Hoflund2021}.

In this work, we study HHG by a non-ideal, slightly astigmatic, driving field. Combining simulations and experimental results, we highlight the impact of a realistic driving field on the wavefront properties of the generated XUV beam, and the consequences for the achievable refocused XUV intensity.
We demonstrate that a high degree of control on the driving field wavefront is necessary to fully remove the induced aberrations of the XUV field. The model presented allows pinpointing the parameter space for the generation of high-quality XUV beams independently of the particular focusing geometry. Therefore, the results presented offer general guidelines for different experimental setups. 

This article is organized as follows: In section~\ref{Sec:ModelDescription}, we describe the model for high-order harmonic generation using astigmatic driving fields and in section~\ref{Sec:Experiment} we present the experimental results. In section~\ref{Sec:Discussion}, we discuss the results and investigate the variation of the XUV beam properties over the parameter space spanned by the driving field astigmatism and beam asymmetry, and we conclude in section~\ref{Sec:Conclusion}.

\section{Model} \label{Sec:ModelDescription} 
\subsection{Astigmatic driving fields}
\begin{figure}[!ht]\centering
\includegraphics[width=13 cm]{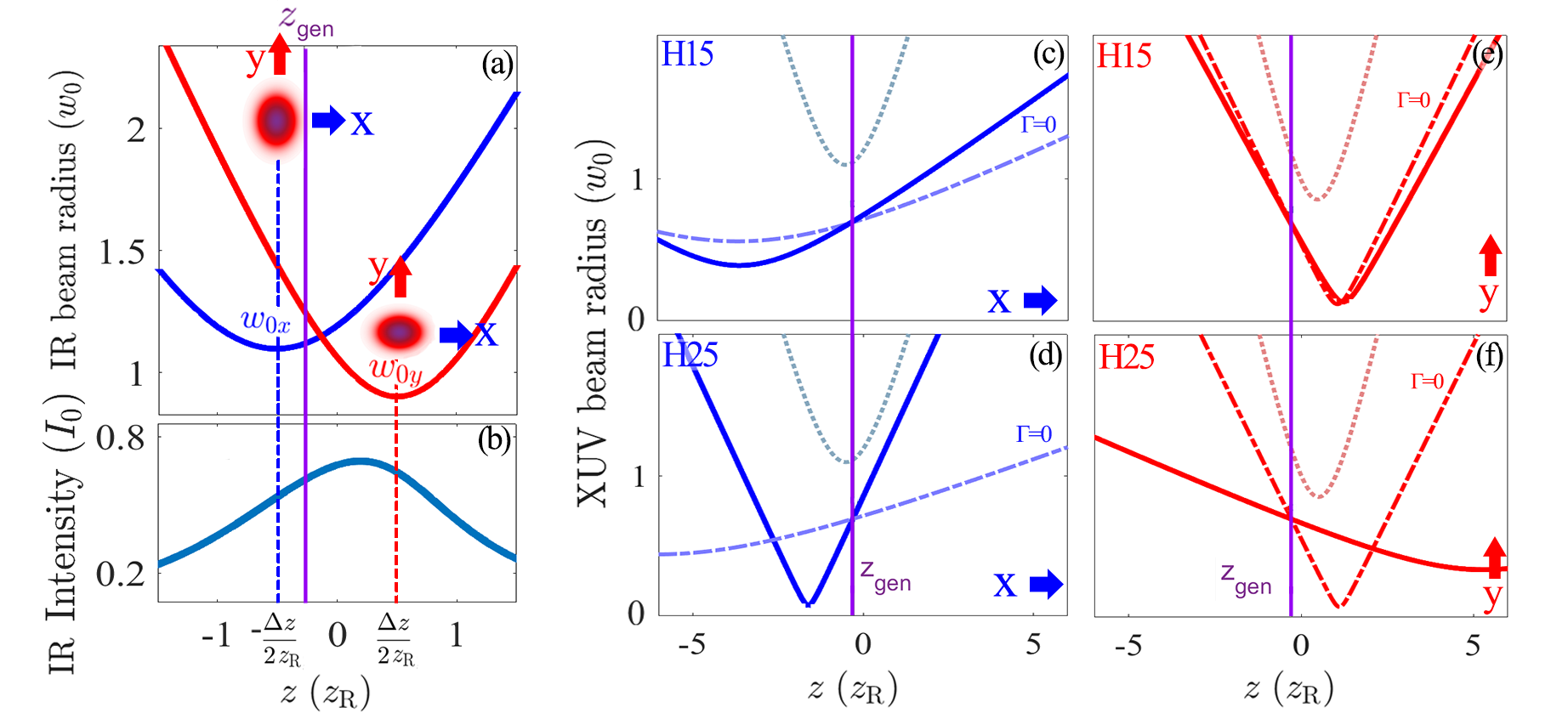}
\caption{(a) Variation of the beam radii of an astigmatic driving field along $x$ (blue solid line) and $y$ (red solid line) as a function of the coordinate $z$ in units of the average Rayleigh length, $z_{\mathrm{R}}$, along the propagation axis. The vertical solid purple line marks the generation plane, $z_\mathrm{gen}$. 
(b) On-axis intensity of the astigmatic driving field as a function of $z$ in units of $I_0$.
(c), (d), Variation of the beam radii along $x$ (blue) for harmonics 15 and 25, respectively, assuming astigmatic Gaussian harmonic beams, i.e. neglecting the higher-order terms. The dashed blue line corresponds to the propagation without consideration of the dipole phase ($\Gamma=0$). The thinner dotted blue line indicates the positions of the $x$ focus of the driving field from panel (a). (e) and (f), same as (c) and (d) for the $y$-direction. 
}
\label{FigModel}
\end{figure}

Let us consider a driving field exhibiting astigmatism, i.e. having different focal positions along two perpendicular axes as well as possibly different beam waists for the two foci. A simple astigmatic Gaussian beam can be modeled as a product of two Gaussian beams~\cite{Kochkina:13}, with the two axes $x,y$ selected to coincide with the principal semi-axes of the beam profile. The spatial phase, $\phi$, and intensity profile, $I$, of an astigmatic beam, propagating along the $z$-axis, can be written as
\begin{align}
\label{eq:PHIG}
\phi(x,y,z)&=\frac{kx^2}{2R_x(z)}+\frac{ky^2}{2R_y(z)},\\
I(x,y,z)&= I_0\frac{w_{0x}w_{0y}}{w_{x}(z)w_{y}(z)}\exp\left(-\frac{2x^2}{w_x^2(z)}-\frac{2y^2}{w_y^2(z)}\right),
\label{eq:IG}
\end{align}
where $k$ is the wave number, and $I_0$ is the maximum peak intensity achievable in the absence of astigmatism. We here focus on the radial dependence of the phase and omit to write the Gouy phase explicitly. 
$R_x(z)$ is the radius of curvature and $w_x(z)$ the beam radius along the $x$-axis, defined as
\begin{align}
R_x(z)=z-z_{0x}+\frac{z_{\mathrm{R}x}^2}{z-z_{0x}},
\hspace{1em} w_x(z)=w_{0x}\sqrt{1+\frac{(z-z_{0x})^2}{z_{\mathrm{R}x}^2}}\label{eq:Wxy},
\end{align}
and similarly for $R_y$ and $w_y$. Here, $z_{\mathrm{R}x}$ ($z_{\mathrm{R}y}$) is the Rayleigh length and $z_{0x}$ ($z_{0y}$) the position of the $x$ ($y$) focus along the $z$ direction. The distance between the two foci, or the astigmatic difference, is denoted $\Delta z=z_{0y}-z_{0x}$, and the average Rayleigh length is $z_{\mathrm{R}}=(z_{\mathrm{R}x}+z_{\mathrm{R}y})/2$.
Fig.~\ref{FigModel} shows the evolution of the beam radii (a) for an astigmatic driving field with $\Delta z= z_\mathrm{R}$ and $w_{0x}/w_{0y} = 1.15/0.9$, as well as the variation of the on-axis intensity (b).
The origin ($z=0$) is defined as the center point of the two foci, which are then located at $z = \pm \Delta z/2$. 
The vertical axis is normalized to the average driving field beam waist, $w_0=(w_{0x}+w_{0y})/2$. 
The ellipses in Fig.~\ref{FigModel}(a) represent the driving field intensity profile at the focus positions, marked by the vertical dashed blue and red lines.

\subsection{The $\gamma$-model for HHG} \label{SubSec:gamma_model}
High harmonic generation is the result of a phase-matched superposition of single atomic responses to an intense IR field (10$^{14}$-10$^{15}$ W/cm$^2$)~\cite{PhysRevLett.90.193901,Heyl_2016}. The XUV radiation produced via HHG by an individual atom has been successfully explained by the strong field approximation (SFA)~\cite{PhysRevA.49.2117}, which exploits an approximate solution of the time-dependent Schrödinger equation. 
The SFA recovers and justifies the semi-classical description of attosecond pulse generation, known as the three-step model~\cite{PhysRevLett.68.3535,PhysRevLett.71.1994,PhysRevLett.70.1599}, which assumes that the electron, after tunnelling from its parent ion, propagates in the continuum, following classical trajectories, before recombining with its ion and emitting an XUV photon at a frequency $\Omega$. The phase of the harmonic radiation is the sum of the phase transferred from the driving field and the so-called dipole phase accumulated during the electron trajectory in the continuum. 

In the three-step model, the variation of the return energy as a function of return time can be well approximated by a straight line in the plateau regime, for each family of electron trajectories~\cite{Guo2018,HampusSpatioTemporal}. This observation leads to an analytical description of the dipole phase, $\Phi_i(\Omega)$, with $i=\left[s,l\right]$ denoting the trajectory (short or long), expressed as 
\begin{equation}
\Phi_i(\Omega)=\alpha_iI+t_{pi}(\Omega-\Omega_p)+\dfrac{\gamma_i}{I}(\Omega-\Omega_p)^2,
\label{eq:Phase-gamma}
\end{equation}
where $\hbar\Omega_p$ is the ionization energy of the atom, $I$ is the local laser intensity, and $\alpha_i$ and $t_{pi}$ are constants that can be numerically obtained within the model, in good agreement with SFA calculations~\cite{Carlstrm2016}. 
In particular, $\alpha_s= 0$ and $\gamma_s=0.22mc/(\alpha\lambda)$ for the short trajectories, whereas $\alpha_l=\alpha\lambda^3/(2\pi m c^3)$ and $\gamma_l=-0.19mc/(\alpha\lambda)$, for the long trajectories;
$\lambda$ is the laser wavelength, $\alpha$ the fine structure constant, $m$ the electron rest mass, and $c$ the speed of light. This analytical model, which we call the $\gamma$-model, shows an excellent agreement with the predictions of the SFA~\cite{Weissenbilder2022}.

In the following, we only discuss the short trajectories, which are typically the dominant contribution to the harmonic yield. The dipole phase affects the spatial properties of the harmonics through its intensity dependence, implying that only the last term in Eq.~\eqref{eq:Phase-gamma} needs to be considered.

By assuming a homogeneous infinitely thin generation medium, we disregard the role of macroscopic phase matching.{ While the macroscopic response can also influence the spatial properties of the harmonics~\cite{Weissenbilder2022}, a study of such effects are beyond the scope of the present work}. In {the considered} scenario, the XUV radiation generated by an IR beam at the generation plane, $z_\text{gen}$, is simply a spatial superposition of individual atomic responses, each driven by a different {local value of} the electric field $E(x,y,z_\text{gen})$. 
As a consequence, following the derivation proposed in~\cite{dacasa2019single, HampusSpatioTemporal}, we can write the phase of the $q$-th harmonic field emitted by short trajectories, $\phi_{q}$, as
\begin{equation}
\phi_{q}(x,y,z_\text{gen})=q\phi(x,y,z_\text{gen})+\dfrac{\gamma_s}{I(x,y,z_\text{gen})}(q\omega-\Omega_p)^2,
\label{eq:GeneralizedGModel}
\end{equation}
where $q\phi$ is the phase transferred from the infrared field and $\omega=\Omega/q$ is the laser frequency.
Furthermore, we assume that the harmonic intensity, $I_q$, varies (on average) according to a power law of the driving field intensity~\cite{Weissenbilder2022,HernndezGarca2015,PhysRevLett.117.163202}, 
\begin{equation}
I_{q}(x,y,z_\text{gen})=[I(x,y,z_\text{gen})]^p, 
\label{eq:iq}
\end{equation}
with $p\approx2.6$ according to ref.~\cite{Weissenbilder2022}.

\subsection{HHG with an astigmatic driving field}

The next step of the derivation consists in replacing $\phi$ and $I$ in Eq.~\eqref{eq:GeneralizedGModel} by their expression for an astigmatic beam [Eqs.~\eqref{eq:PHIG} and \eqref{eq:IG}], making a Taylor expansion of the intensity around $x=y=0$, and keeping only the terms proportional to $x^2$ and $y^2$. We then obtain the phase up to the second order in $x$ and $y$
\begin{equation}
\phi_{q}^{(2)}(x,y,z_\text{gen})=\frac{qkx^2}{2R_{qx}(z_\text{gen})}+\frac{qky^2}{2R_{qy}(z_\text{gen})},
\label{eq:phasetoto}
\end{equation}
with
\begin{align}
\label{eq:radiusRq}
\frac{1}{R_{qx}(z_\text{gen})}&=\frac{1}{R_x(z_\text{gen})}+\frac{4\Gamma c}{q\omega w_x^2(z_\text{gen})},
&\Gamma=\frac{\gamma_s(q\omega-\Omega_p)^2 w_{x}(z_\text{gen})w_{y}(z_\text{gen})}{I_0 w_{x0}w_{y0}}.
\end{align}
and similarly for $R_{qy}(z_\text{gen})$. The knowledge of $R_{qx}$, $R_{qy}$, as well as the radii $w_{qx}(z_\text{gen})=w_x(z_\text{gen})/\sqrt{p}$ and $w_{qy}(z_\text{gen})=w_y(z_\text{gen})/\sqrt{p}$ [from Eq.~\eqref{eq:iq}] in the generation plane allows an analytical description of the generated harmonic beams as astigmatic Gaussian beams. The transfer of the astigmatism of the driving field to the harmonics is, however, non-trivial because of the dipole phase contribution in the second term of Eq~\eqref{eq:radiusRq}. The calculated variations of the beam radii are shown in Fig.~\ref{FigModel} for the 15th (c,e) and 25th (d,f) harmonics generated in argon for the $x$ (c,d) and $y$ (e,f) axes, respectively.
The dashed lines show the trivial transfer of phase from the driving field, corresponding to $\Gamma = 0$, while the solid lines include the dipole phase contribution to the radii of curvature.

Since the position of the $x$ driving field focus is before the generation plane, the harmonics are generated as diverging beams with virtual foci for the $x$-direction [Fig.~\ref{FigModel} (c,d)]. 
In this case, the dipole phase contribution leads to higher divergence for the higher-order harmonics, due to the $q$-dependence of $\Gamma$.
The position of the $y$ driving field focus is well after the generation plane, and the harmonics are generated as converging beams with real foci along the $y$-axis [Fig.~\ref{FigModel} (e,f)]. In this case, the dipole phase contribution counteracts the trivial phase transfer, leading to a smaller convergence for higher-order harmonics and, thus, also to a smaller divergence after the real focus. As has been highlighted in earlier studies~\cite{HampusSpatioTemporal}, but not shown here, in this case, the dipole phase contribution can also overcome the trivial phase contribution, leading to virtual foci of the harmonics, even for a converging driving field.
To compare the predictions of our model with the experimental results (see Fig.~\ref{fig:Figure_1_schematic}), we sum the square of the harmonic spatial profiles at the distance of observation. Each profile is multiplied by a coefficient reflecting the variation of the experimentally measured harmonic yields. 

\subsection{Higher-order contributions} \label{SubSec:Higher-order_cont}
Our model can be refined by including higher-order terms of the Taylor expansion of the dipole phase. Eq.~\eqref{eq:phasetoto} becomes
\begin{equation}
\phi_{q}^{(2N)}(x,y,z_\text{gen})=\phi_{q}^{(2)}(x,y,z_\text{gen})+\Gamma\sum_{n=2}^N\frac{1}{n!}\left(\frac{2x^2}{w_x^2}+\frac{2y^2}{w_y^2}\right)^n.
\label{eq:phasetotoh}
\end{equation}
The higher-order terms, with $2 \leq n \leq N$, introduce higher-order spatial aberrations, whose impact will be discussed in section~\ref{Sec:exp_ho}.
In this case, to compare the predictions of our model with the experimental results, we numerically calculate diffraction in the Fraunhofer regime and sum over the different harmonics of the spectrum in the XUV far-field
\begin{equation}
I_{XUV}(\theta_x,\theta_y)=\sum_q\Big|\iint \sqrt{I_q(x,y,z_\text{gen})}e^{-i\phi_q(x,y,z_\text{gen})} e^{i qk x\theta_x}e^{i qk y\theta_y}dxdy\Big|^2,
\label{eq:AllHarmonicsSummed}
\end{equation}
$\theta_x$, $\theta_y$ denoting the emission angles in the $x$ and $y$ axes. For the simulations presented below, we use $N=4$.

\section{Experiment} \label{Sec:Experiment}
\begin{figure}[ht]\centering
\hspace*{-0.5cm}
\includegraphics[width=14 cm]{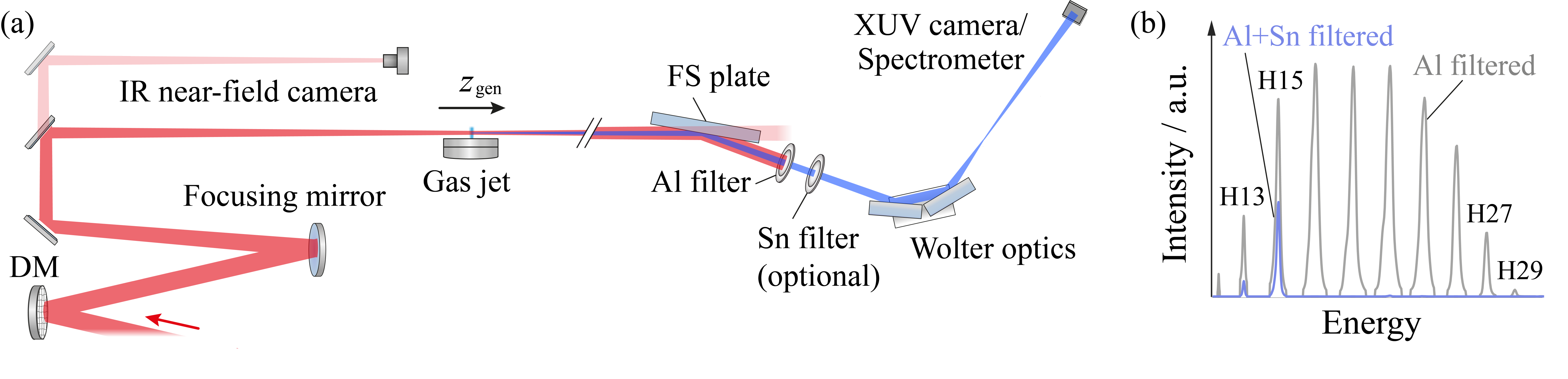}
 \hspace*{+0.5cm}
\caption{ (a) Schematic of the Intense XUV Beamline providing high-flux high-order harmonics by HHG in an argon gas jet and a loose focusing geometry. Different filtering can be applied before refocusing and recording the XUV spatial profile on a CCD camera, or the XUV spectrum using a grating spectrometer. (b) XUV spectrum using an aluminum filter (gray) and an aluminum plus a Sn filter (blue).}
\label{fig:experiment}
\end{figure}

\subsection{Experimental setup}
The experimental setup is shown in Figure~\ref{fig:experiment}. A terawatt laser system provides infrared (IR) pulses with 40 fs duration, a central wavelength of 800~nm, and up to $\unit[45]{mJ}$ pulse energy at a repetition rate of 10 Hz. The deformable mirror (DM) is a membrane mirror featuring 32~piezoelectric actuators on its backside. It is used to shape the wavefront of the pulses by adapting its surface structure in a controllable manner. A dielectric focusing mirror ($f = \unit[8.7]{m}$) focuses the beam in the generation gas (argon). In addition, varying the DM curvature allows moving the focus with respect to the generation gas. A leak through the last mirror before the gas cell is used to image the near-field (focus) of the driving beam.

High-order harmonics are generated in a pulsed argon gas jet. The remaining IR is reduced by transmission through a fused silica (FS) plate while the XUV radiation is reflected. Filtering with a 200 nm thick aluminum film takes out the remaining IR and selects harmonics of order higher than 13. In addition, filtering with a Sn film allows us to select predominantly the 15th harmonic. 

The beam is then refocused by two toroidal mirrors in a Wolter configuration with a de-magnification factor of 35. A back-illuminated XUV CCD camera (Andor iKon-L) monitors the XUV far-field profile. By moving in a concave Hitachi grating, the XUV spectrum can be recorded using an multi-channel plate (MCP)-phosphor screen assembly imaged by a CCD camera. Harmonic spectra obtained with and without the Sn filter are shown in Fig.~\ref{fig:experiment} (b).

\begin{figure}[ht]\centering
\includegraphics[width=12 cm]{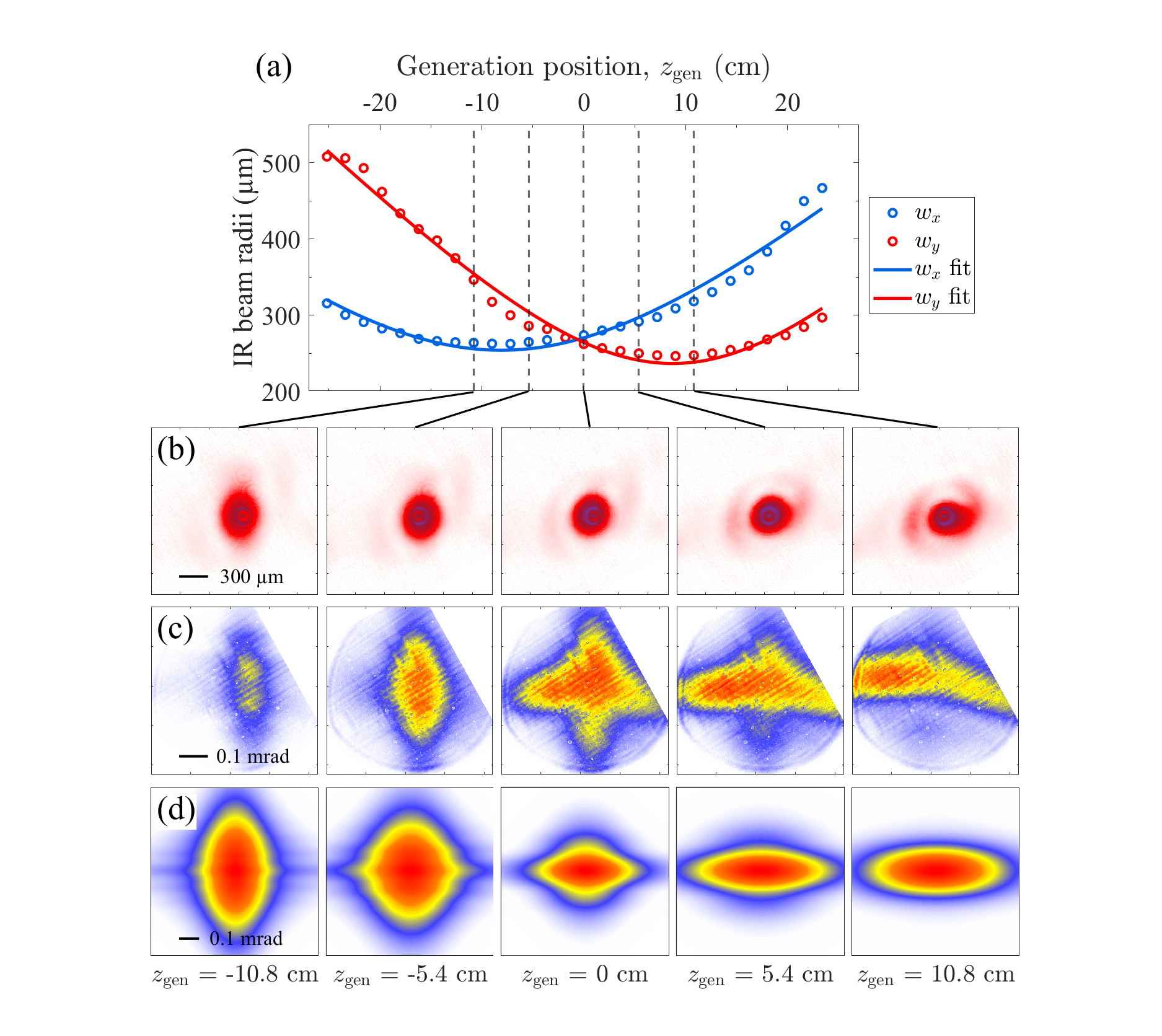}
\caption{(a) Extracted IR beam radii in the $x$ (blue) and $y$ (red) directions. The open circles show the experimental data, while the solid lines are obtained by fitting the beam radius of a Gaussian beam to the experiment; (b) IR focus images at different positions along the $z$ axis.
(c) XUV far-field profiles for generation at the same positions; (d) Simulations of the XUV far-field profiles, including the higher-order terms, using the fit from figure (a) as simulation input. The experimental divergence indicated in figure (c) is calculated as the far-field divergence of the XUV before refocusing by the Wolter optics.}
\label{fig:Exp_coll_uncorrected}
\end{figure}

\subsection{Experimental results using an aberrated driving field}
Figure~\ref{fig:Exp_coll_uncorrected} shows the IR and XUV intensity profiles for different generation positions, when no aberration correction is applied to the DM. In order to simplify the presentation, the images are rotated so the $x$ and $y$ axes are horizontal and vertical, respectively. The IR beam focus is slightly astigmatic with different foci along the $x$ and $y$ axes, as shown in Fig.~\ref{fig:Exp_coll_uncorrected} (a) and (b). The Rayleigh lengths extracted from the fits are $z_{\mathrm{R}x} = \unit[22.3]{cm}$ respectively $z_{\mathrm{R}y} = \unit[17.5]{cm}$. The distance between the foci is $\unit[16.8]{cm}$.

The corresponding XUV far-field profiles are shown in Fig.~\ref{fig:Exp_coll_uncorrected} (c). The images are cut by a circle due to propagation through an aperture, and a straight line due to the edge of the camera chip. The measured profiles exhibit strong spatial distortion, varying from vertical to horizontal with cross-like patterns at the intermediate generation positions. 

Fig.~\ref{fig:Exp_coll_uncorrected} (d) shows the results of simulations performed using the model presented in section~\ref{Sec:ModelDescription}, including the higher-order contributions. The parameters used for the simulations are chosen to mimic the experimental conditions. We use, in particular, the fitted IR beam radii indicated by the solid lines in Fig.~\ref{fig:Exp_coll_uncorrected} (a).
The simulations agree well with the experimental results and further allow us to understand the observed patterns, as discussed in the following.

\begin{figure}[ht]\centering
\includegraphics[width=10 cm]{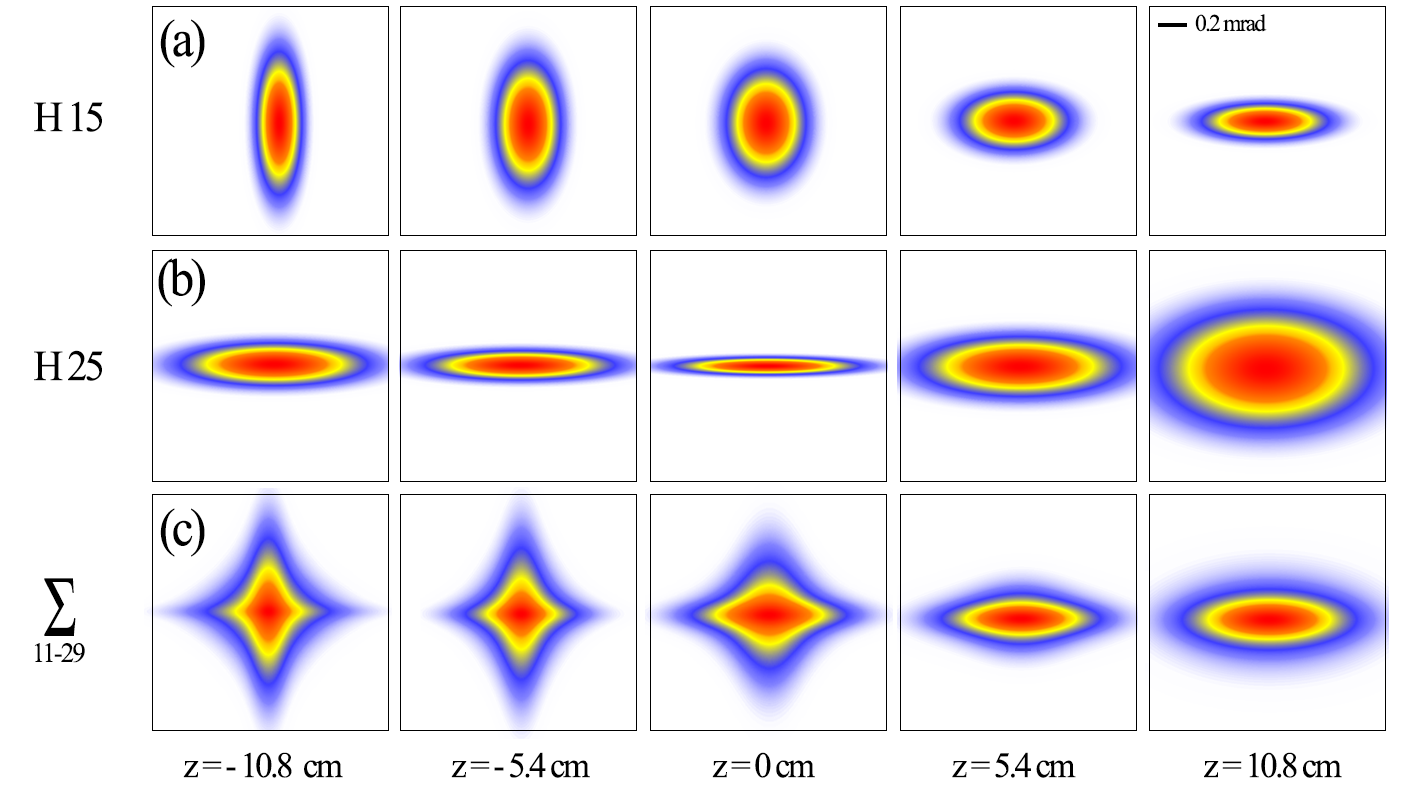}
\caption{Simulated XUV far-field profiles for harmonic orders 15 (a) and 25 (b), as well as the sum of all contributing harmonic orders from 11 to 25 (c), taking into account their relative spectral weights. The columns correspond to the same generation positions as in Fig.~\ref{fig:Exp_coll_uncorrected}, indicated below the figure. The simulations were performed without inclusion of the higher-order terms.}
\label{fig:comp-simulation-exp}
\end{figure}

\subsection{The source of the spatial distortions}

Figure~\ref{fig:comp-simulation-exp} presents the XUV far-field profiles for the same generation positions as in Fig.~\ref{fig:Exp_coll_uncorrected}, from simulations performed using the model presented in section~\ref{Sec:ModelDescription}, excluding the higher-order contributions. We show harmonics 15 (a) and 25 (b) as well as the sum of all harmonics from 11 to 29 (c), taking the relative intensities into account as described above. A figure including the individual far-field profiles of all harmonics is included in the SI.

For harmonic 15, the profile changes from vertically elongated to horizontal when the generation position is moved later. For higher harmonic orders, the same change from vertical to horizontal takes place, but the transition occurs at earlier generation positions (see the SI). For harmonic 25 the profile is horizontal at all generation positions.

This behaviour can be understood based on the earlier discussion on the divergence of harmonics 15 and 25 in Fig.~\ref{FigModel}. When the generation position is moved from -10.8 to 10.8 cm, it moves between the focus in the $x$-direction and the focus in the $y$-direction, which means that the IR wavefront contribution to the XUV wavefront is always diverging in the $x$-direction and converging in the $y$-direction.
At $z_\text{gen} = -10.8$~cm, without the dipole phase contribution to the XUV wavefront curvature, the XUV wavefront would be almost flat in the $x$-direction, but converging in the $y$-direction, leading to a larger far-field divergence in the $y$-direction, and a vertically elongated far-field profile. Including the dipole phase contribution, this is still the case for harmonic 15, where the dipole contribution is small. For harmonic 25, however, where the dipole contribution is larger, the divergence will increase in the $x$-direction and decrease in the $y$-direction (as discussed in section~\ref{Sec:ModelDescription}), leading to a horizontally elongated far-field profile.
At $z_\text{gen} = 10.8$~cm, without the dipole phase contribution to the XUV wavefront curvature, the XUV wavefront would be divergent in the $x$-direction, but almost flat in the $y$-direction, leading to a larger far-field divergence in the $x$-direction, and a horizontally elongated far-field profile. Including the dipole phase contribution, this is still the case for all harmonics, since the dipole phase contribution will increase the divergence both along $x$ and $y$.

Comparing these results with those obtained including the higher-order contributions in Fig.~\ref{fig:Exp_coll_uncorrected} (d), we see that the model without the higher-order contributions can explain the cross-like patterns observed in the far-field profile as a superposition of individual harmonics that are either horizontally or vertically elongated. However, the inclusion of the higher-order terms leads to additional structures, in particular when the XUV wavefront is almost flat, further distorting the profiles, as discussed in the following.

\begin{figure}[ht!]\centering 
\includegraphics[width= 10 cm]{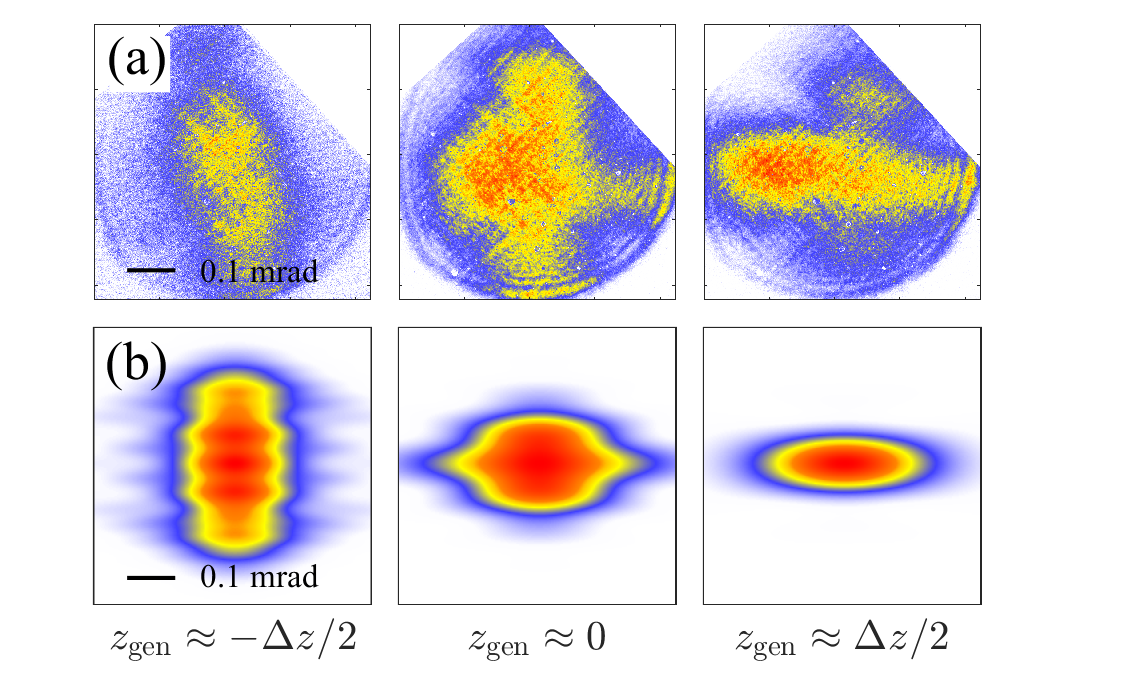}
\caption{
Measured (a), and simulated (b) far-field profiles of the 15th harmonic for different generation positions.
}
\label{fig:SnAlongPropagation}
\end{figure}

\subsection{Spatial profiles of individual harmonics} 
\label{Sec:exp_ho}

To investigate the effect of the higher-order terms of the phase transfer introduced in Eq.~\eqref{eq:phasetotoh}, until now hidden in the superposition of different harmonic orders, we isolate a single harmonic by inserting a $\unit[200]{nm}$ Sn filter into the beam path, resulting in predominantly the 15th harmonic. Fig.~\ref{fig:experiment} (b) shows the harmonic spectrum with, and without the Sn filter. 
 
Figure~\ref{fig:SnAlongPropagation} shows the measured (a) and simulated (b) 15th harmonic spatial profiles. The experimental conditions are different from those in Fig.~\ref{fig:Exp_coll_uncorrected}.
When the generation position is close to the focus in the $x$-direction, the 15th harmonic far-field profile is vertical. When the generation position is closer to the focus in the $y$-direction, the 15th harmonic profile presents a cross-like pattern. The simulations, which include the higher-order contributions, show similar behavior.
These features are normally hidden under the superposition of several harmonics and have a different physical origin. They are due to interference between different parts of the non-parabolic wavefront having the same propagation direction.

\section{Discussion} \label{Sec:Discussion}
As shown in Fig.~\ref{FigModel} (c-f), and discussed earlier, each harmonic has different beam waist sizes and beam waist positions for the $x$ and $y$ foci. As a consequence, the beam divergence of the $q$th harmonic, $\theta_{qx}=\lim_{z\rightarrow \infty}(w_{qx}(z)/z)$ in the $x$-direction and similarly $\theta_{qy}$ in the $y$-direction will significantly change with the generation position and the aberrations of the driving field, impacting the properties of the XUV radiation. 
While the experimental results were obtained for a given set of parameters for the astigmatism of the driving field, we will now turn to a more systematic study of the variation of the XUV properties with the driving field astigmatism.

In a typical experiment, the most direct observable for the XUV beam quality is the far-field profile of the harmonics, while what is typically important for applications is the quality of the refocused XUV pulses, which are strongly dependent on aberrations.
To represent such a direct experimental observable describing the far-field properties of the $q$th harmonic, we introduce an eccentricity parameter $\Theta_{qxy}$,
\begin{equation}
\Theta_{qxy}=\frac{\theta_{qy}-\theta_{qx}}{\theta_{qy}+\theta_{qx}},
\label{eq:Asymm}
\end{equation}
which one in a typical experiment would strive to minimize in magnitude in order to achieve a circularly symmetric XUV beam profile.

\begin{figure}[ht]\centering
\includegraphics[width=9 cm]{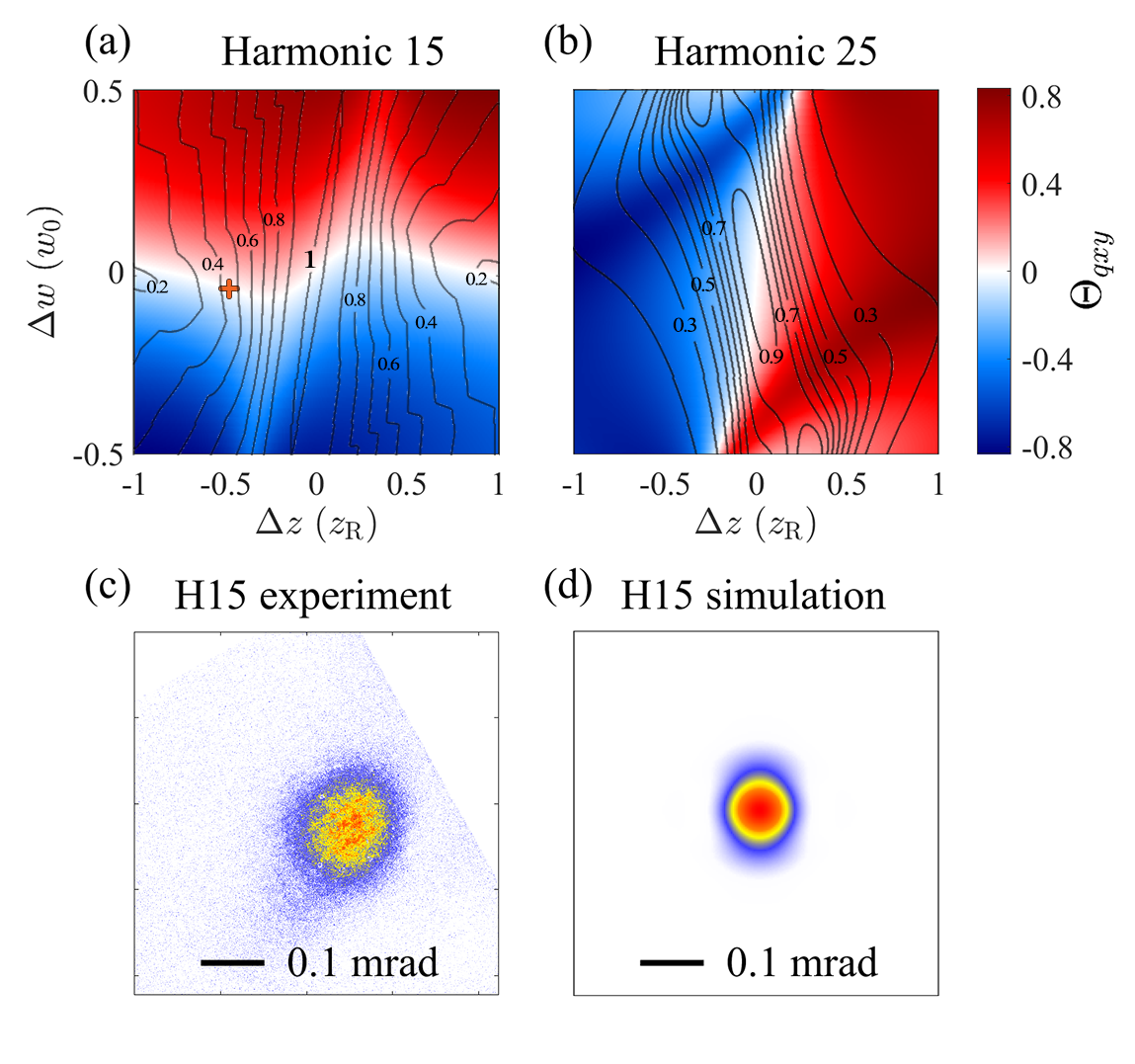}
\caption{Variation of the eccentricity parameter, $\Theta_{qxy}$, for harmonic 15 (a) and 25 (b) as a function of the astigmatism ($\Delta z/z_\text{R}$) and beam waist radii difference ($\Delta w/w_0$) of the driving field. Black contour lines indicate the Strehl-ratio. (c) experimentally measured and (d) simulated far-field profile of the 15th harmonic at the point marked by the red cross in panel (a). The generation position of the simulation coincides with the one used for the experimental profile, $z_\text{gen}=-0.2z_\text{R}$.}
\label{fig:PracticalAdvice}
\end{figure}

In Fig.~\ref{fig:PracticalAdvice} we plot this eccentricity parameter for harmonics 15 and 25, as a function of the driving field astigmatism, defined through the astigmatic distance $\Delta z$, in units of $z_\mathrm{R}$, and the waist radii difference $\Delta w=w_{0x}-w_{0y}$, in units of $w_0$.
For the plots, we use our analytical model, without higher-order contributions, for one generation plane, $z_\text{gen}=-0.2\;z_\text{R}$, and one peak intensity $I_0=2\times10^{14}$~W/cm$^2$, close to the experimental conditions discussed above.

When $\Theta_{qxy}>0$ (red regions), the harmonic far-field profile is vertically elongated, and when $\Theta_{qxy}<0$, (blue regions), it is horizontally elongated. Both harmonics 15 and 25 exhibit regions where the eccentricity parameter is close to zero (white), characteristic of a circularly symmetric profile in the far field, although the driving field and thus the harmonics are astigmatic. While the profile of the 15th harmonic is mostly affected by changing the relative beam waist radii of the driving field, that of the 25th harmonic is strongly affected by the astigmatism of the driving field. 

The fact that the far-field eccentricity of the harmonic beam can be zero despite the harmonic being astigmatic can be understood as a balance between the astigmatic distance and the difference in beam waist radii in $x$ and $y$ of the driving field, which through the intensity-dependent dipole phase contribution leads to the same far-field divergence along $x$ and $y$.
However, that the harmonics are astigmatic might compromise their use in experiments where they are typically refocused. To quantify this, the black contour lines in panels (a) and (b) of Fig.~\ref{fig:PracticalAdvice} indicate how the Strehl ratio of the single harmonic beam varies with the astigmatism and beam waist radii difference of the driving field. In this case, the Strehl ratio is the ratio between the on-axis peak intensity for the astigmatic harmonic beam, and the on-axis peak intensity for the stigmatic harmonic beam, generated by a stigmatic driving field with $\Delta z = \Delta w = 0$.

As an example, the lower row of Fig.~\ref{fig:PracticalAdvice} shows the experimentally measured (c) and simulated (d) far-field profiles of the 15th harmonic at the point marked by the red cross in panel (a). In this case, the far-field profile of the harmonic is not eccentric, while the harmonic astigmatism leads to a Strehl ratio of $\approx$ 0.5. Such a sub-optimal peak intensity of the refocused XUV beam can be detrimental for experiments requiring high XUV intensities, such as studies of non-linear effects~\cite{PhysRevA.93.061402} and pump-probe studies. 

While the Strehl ratio for a single harmonic decreases away from the point $\Delta z = \Delta w = 0$, it is interesting to note that reasonably high Strehl ratios can be maintained for quite large differences in beam waist radii by a small compensation by the driving field astigmatism. 
In these regions, the harmonic beam is almost stigmatic, but with different divergence in $x$ and $y$, leading to an elliptical focal spot. This might be sub-optimal for certain applications, such as imaging and metrology, despite the high Strehl ratio~\cite{Miao15}.

For a superposition of several harmonics, as required to achieve attosecond pulse durations, the variation of the astigmatism with harmonic order will further lead to sub-optimal spatial overlap of the harmonics, impacting also the temporal qualities of the pulses~\cite{Hoflund2021}.

\begin{figure}[ht!]\centering
\includegraphics[width=12 cm]{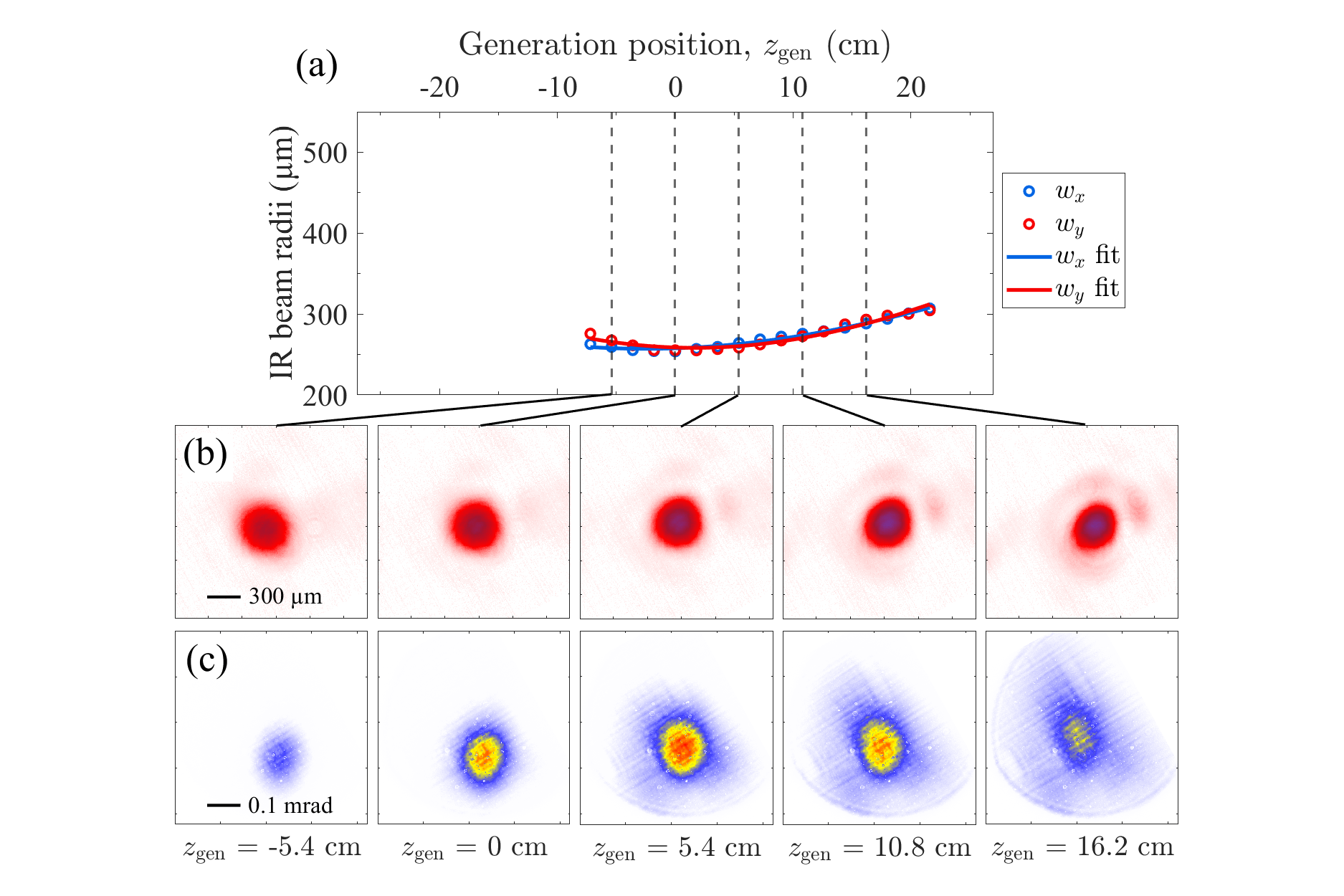}
\caption{ (a) Extracted IR beam waists in the $x$ (blue) and $y$ (red) directions for an aberration-corrected beam; (b) IR focus images at different positions along the $z$-axis. (c) XUV far-field profiles resulting from generation at the same positions.
}
\label{fig:exp-corrected}
\end{figure}

From this study, one can conclude that, in order to ensure optimal focusability of XUV pulses from HHG, it is crucial to generate them with a non-aberrated, stigmatic driving field. As a demonstration, our experimental setup allows us to correct the aberrations of the driving field by acting on the settings of the deformable mirror. Figure~\ref{fig:exp-corrected} (a) presents the variation of the beam waists of the corrected IR beam along the propagation direction $z$, for the $x$- and $y$ axes. In Fig.~\ref{fig:exp-corrected} (b), we show the evolution of the IR beam close to the focus. Figure~\ref{fig:exp-corrected} (c) presents the XUV far-field profiles at the corresponding generation positions. 

For the stigmatic driving field, the XUV far-field profiles are symmetric and only weakly dependent of the generation position (c.f. Fig.~\ref{fig:Exp_coll_uncorrected}). While the IR spatial profiles are similar in Fig.~\ref{fig:exp-corrected} (b) and~\ref{fig:Exp_coll_uncorrected} (b), the XUV profiles are strongly improved by correcting the small remaining aberrations of the driving field. This shows that even a small degree of astigmatism in the IR can cause strong distortions of the XUV far-field profiles. Aberration correction of the driving field can dramatically improve the XUV beam quality.

\section{Conclusion} \label{Sec:Conclusion}
In this work, we have studied the influence of astigmatic driving fields on the far-field profiles of high-order harmonics. We present an extension of the model presented in~\cite{HampusSpatioTemporal}, taking into account the influence of an astigmatic driving field wavefront, and study the eccentricity of the far-field profiles of single harmonics as well as of the superposition of several harmonics. 
We show that the far-field eccentricity depends on the astigmatism of the driving field, the generation position, the peak intensity, and the harmonic order. By taking into account the interplay between the dipole phase contribution to the wavefront of each harmonic as well as the inherited driving wavefront profile, different harmonic orders exhibit astigmatism with two different source positions along the propagation axis of the driving field. 

As observed in our experiments and reproduced by the simulations, we show that a cross-like pattern can appear in the broadband harmonic spatial profile, resulting from a superposition of individual harmonic far-field profiles which are vertically or horizontally elongated.
We demonstrate that such features can also appear in the spatial profiles of the individual harmonics, which can be explained by taking into account high-order terms in the theoretical model. 

The model allows us to explore the phase space and give information on the best generation conditions to minimize the XUV aberrations. We show that non-eccentric XUV beams can be generated also by slightly aberrated driving fields due to a balance between the astigmatic distance and the difference in beam waist radii between the principal semi-axes of the astigmatism. However, in such cases, the XUV beams are astigmatic, and after refocusing the maximum achievable intensity is not reached, as quantified by the calculated Strehl ratio. We also demonstrate that, through a similar balance, stigmatic harmonic beams with Strehl ratios close to one can be generated. In such cases, however, the refocusing results in elliptical focal spots, compromising the utility of such beams in e.g. imaging and metrology applications.

By means of a deformable mirror, we experimentally adjusted the driving field astigmatism and eccentricity, obtaining aberration-free XUV profile for all generation positions. Our observations indicate that fine-tuning the driving wavefront is essential for controlling the spatial quality of the harmonics. 

In conclusion, high-quality XUV beams reaching the highest peak intensities and the smallest focal spot sizes can only be generated using aberration-free driving fields, underlining the importance of proper 
correction of the driving wavefront. The conclusions can be further extended to different generation positions, driving field intensities and central wavelengths. It is interesting to remark, as shown in Eq.~\eqref{eq:radiusRq}, that as the XUV aberration patterns depend on $\Gamma\propto q^2/I$, upon increasing the intensity the cross pattern will shift toward higher photon energies (higher harmonic order, $q$).
With the spatial qualities of XUV radiation from high-order harmonic generation gaining importance in a number of scientific and industrial applications, further work on the topic should include the effects of higher-order driving field aberrations, like coma and spherical aberration, on the HHG process.
\newpage
\section*{Disclosures}
The authors declare no conflicts of interest regarding the publication of this article.

\section*{Data Availability}
 Data underlying the results presented in this paper (in the form of images, numeric data sets) are not publicly available at this time but may be obtained from the authors upon reasonable request.
 
\section*{Funding}
The authors acknowledge support from the Swedish Research Council (
2013-8185, 
2021-04691,%
2017-04106, 
2021-05992 
), the European Research Council (advanced grant QPAP, 884900) and the Knut and Alice Wallenberg Foundation. A.L. is partly supported by the Wallenberg Center for Quantum Technology (WACQT) funded by the Knut and Alice Wallenberg foundation. M.P. and V.P. acknowledge the support of the Helmholtz Foundation through the Helmholtz-Lund International Graduate School (HELIOS, HIRS-0018).
\newpage
\section*{Supplementary Material}
\begin{figure}[!ht]\centering
\includegraphics[width=11.5 cm]{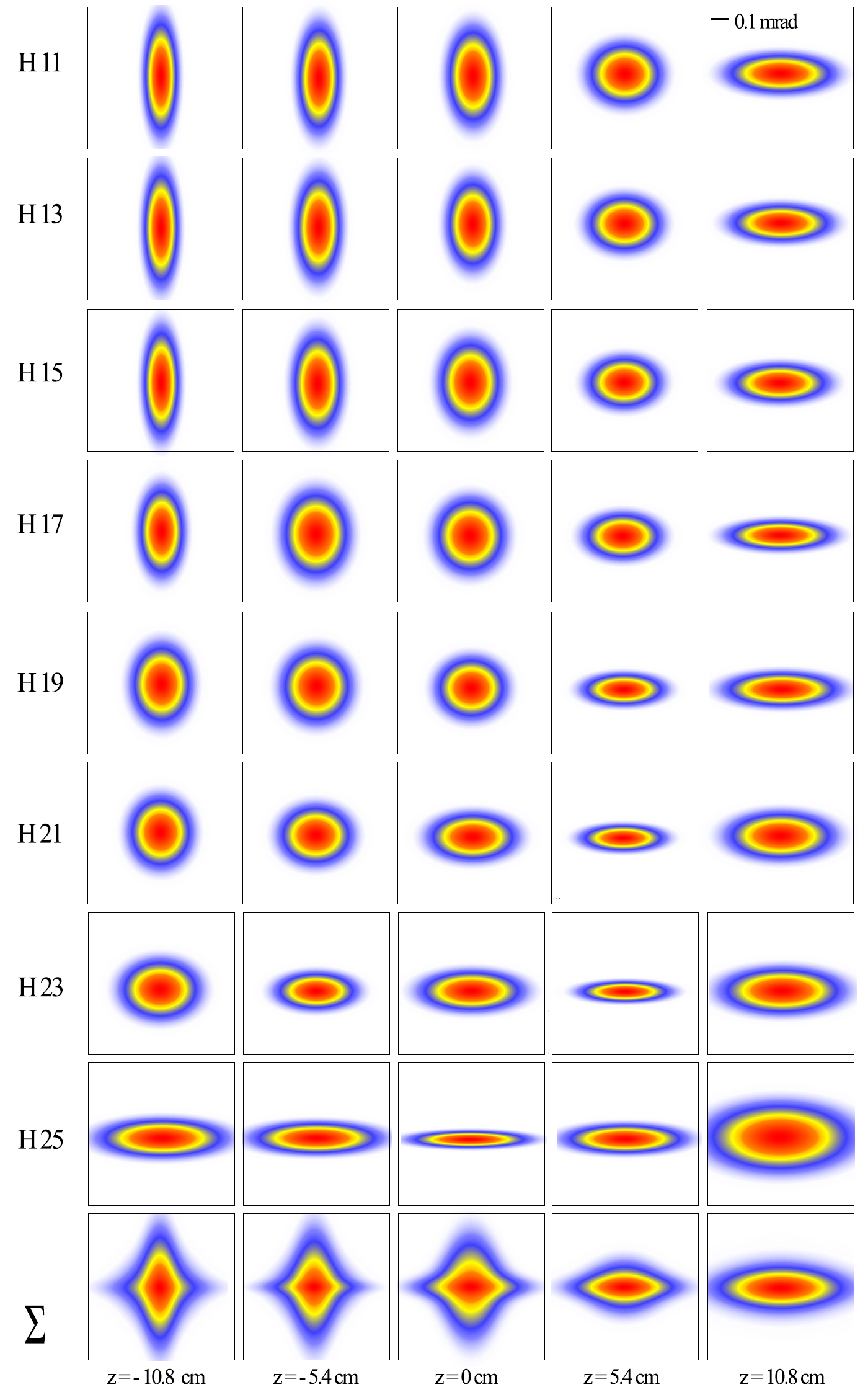}
\caption{From top to bottom: harmonic orders 11 to 25 as well as the sum of harmonics (bottom line), taking into account their relative spectral weights.}
\label{fig:EXTRA}
\end{figure}
\newpage
\newpage
\pagebreak
\subsection*{Author Contributions} 
M.P, F.V., E.A., V.P. conducted the experiments. 
P.E.-J, J.P., C.A. designed the experiment. 
M.P., F.V., E.A. did the data analysis and F.V did the simulations based on theory that P.S., C.A., A.L. developed.
M.P., F.V., E.A., V.P., P.E.-J. and A.L. wrote the article, with feedback from all the authors. 

\section*{Supplementary Materials}
Fig. \ref{fig:EXTRA}. Simulated XUV far-field profiles for harmonic orders 15 to 25 , as well as the sum of all contributing harmonic orders from 11 to 25.
\newpage
\pagebreak
\printbibliography
\end{document}